\documentclass[aps,prl,twocolumn,showpacs,final,a4paper,superscriptaddress]{revtex4-1}

\usepackage{graphicx}
\usepackage{dcolumn}
\usepackage{bm}
\usepackage{color}

\usepackage{amsmath}
\usepackage{ulem}

\begin{document}

\title{Optical signatures of dynamical excitonic condensates}

\author{Alexander Osterkorn}
\email{alexander.osterkorn@ijs.si}
\affiliation{Jo{\v z}ef Stefan Institute, SI-1000, Ljubljana, Slovenia}

\author{Yuta Murakami}
\affiliation{Center for Emergent Matter Science, RIKEN, Wako, Saitama 351-0198, Japan}

\author{Tatsuya Kaneko}
\affiliation{Department of Physics, Osaka University, Toyonaka, Osaka 560-0043, Japan}

\author{Zhiyuan Sun}
\affiliation{State Key Laboratory of Low-Dimensional Quantum Physics and Department of Physics, Tsinghua University, Beijing 100084, P. R. China}

\author{Andrew J. Millis}
\affiliation{Department of Physics, Columbia University, 538 West 120th Street, New York, New York 10027}
\affiliation{Center for Computational Quantum Physics, Flatiron Institute, 162 5th Avenue, New York, NY, 10010}

\author{Denis Gole{\v z}}
\affiliation{Jo{\v z}ef Stefan Institute, SI-1000, Ljubljana, Slovenia}
\affiliation{Faculty of Mathematics and Physics, University of Ljubljana, 1000
Ljubljana, Slovenia}

\begin{abstract}
We theoretically study dynamical excitonic condensates occurring in bilayers with an imposed chemical potential difference and in photodoped semiconductors.
We show that optical spectroscopy can experimentally identify phase-trapped and phase-delocalized dynamical regimes of condensation.
In the weak-bias regime, the trapped dynamics of the order parameter's phase lead to an in-gap absorption line at a frequency almost independent of the bias voltage, while for larger biases, the frequency of the spectral feature increases approximately linearly with bias.
In both cases there is a pronounced second harmonic response.
Close to the transition between the trapped and freely oscillating states, we find a strong response upon application of a weak electric probe field and compare the results to those found in a minimal model description for the dynamics of the order parameter's phase and analyze the limitations of the latter.
\end{abstract}

\maketitle

\paragraph{Introduction.}
Excitons, due to their bosonic nature, can undergo condensation akin to the condensation of Cooper pairs in superconductors
~\cite{jerome1967,keldysh1968,Balatsky2004,Eisenstein2004,Eisenstein2014}. The typically light exciton mass implies the possibility of coherence at high temperatures.
While several materials were proposed to host excitonic condensation in the ground state~\cite{Littlewood2004,wakisaka2009,seki2014,monney2009}, progress in the fabrication of heterostructures~\cite{Geim2013,Novoselov2016,Jin2018} and in  ultrafast techniques~\cite{pareek2024,boschini2024,reutzel2024} have opened the possibility of stabilizing \emph{dynamical excitonic condensates}, i.e. nonequilibrium quantum states where excitons condense in a coherent macroscopic state.
In these, the excitonic population is maintained by a nonequilibrium process involving supplying electrons and holes either by optical excitation~\cite{murotani2019,Perfetto2019,Perfetto2020,Schmitt2022,reutzel2024,Bange2024} or by separately contacting gates to each layer in bilayer structures with interlayer excitons~\cite{zhu1995,littlewood1996,Szymanska2003,Xie2018,wang2019,Ma2021,nguyen2023,qi2023perfect}.
In the latter setup, hybridization of inter-valley excitons with opposing dipoles was recently reported~\cite{Liu2024}.
In either case, the nonequilibrium stabilization of the excitonic phase implies an interlayer chemical potential difference that may drive a time dependence of the condensate.

One recurring problem in the field is how to experimentally detect the coherence of dynamical excitonic condensation.
Recent observations of photo-induced absorption edge enhancement in time-resolved optical spectroscopy~\cite{murotani2019} and fine structure in the angle-resolved photo-emission~\cite{pareek2024} were interpreted as signatures of dynamical condensation.
Dynamical excitonic condensates can host characteristic phenomena like excitonic fluids~\cite{Ma2021,wang2019}, perfect Coulomb drag~\cite{Nandi2012,nguyen2023,qi2023perfect} and coherent photoluminescence~\cite{Ivanov1999,butov2002}.
While measuring transport properties using Coulomb drag is challenging, recent studies claim to observe the perfect Coulomb drag~\cite{nguyen2023}; however, without superfluidity~\cite{qi2023perfect}.

Several theoretical studies analyzed the properties of dynamical excitonic condensates assuming perfect separation between layers, leading to effects like the bandgap renormalization due to injected charge carriers~\cite{Xie2018,zeng2020}, oscillating polarization~(photo-current) with a pump tunable frequency~\cite{Perfetto2019,Murakami2020a} as well as dynamical instabilities of the excitonic condensate~\cite{Hanai2016,hanai2017}.
However, the approximation of perfect decoupling is not always physically reasonable, and weak tunnelling will break the $U(1)$ order parameter symmetry, which can induce finite interlayer resistance in Coulomb drag experiments~\cite{qi2023perfect} and interesting dynamical effects, including the AC Josephson effect~\cite{Sun2021}, tunability between dark and bright excitons~\cite{Sun2023} and leads to the exciton-driven bandgap renormalization~\cite{zeng2024}.
Bright interlayer excitons are particularly interesting since they provide in-plane electromagnetic polarization, which acts as an antenna for out-of-plane emission or absorption of photons.

Therefore, it would be highly desirable to have direct experimental signatures of dynamical exciton condensates similar to early interference measurements of photoluminescence in heterostructures within the Hall regime~\cite{butov2002}.
In this work, we show that the optical response of dynamical excitonic condensate carries important information about the dynamics of the order parameter's phase exhibiting a sharp transition between the trapped and the freely rotating oscillations with an increasing number of excited charge carriers.
We use time-dependent dynamical mean-field theory (tdDMFT) to evaluate the optical response of the system,
as a function of non-equilibrium drive in the trapped and deconfined phases and in particular in the vicinity of the transition.
The general features of this instability can be understood within an effective Landau-Ginzburg theory of the order parameter phase; however, microscopic dynamics show qualitatively different dynamics close to the instability.

\begin{figure}
\centering
\includegraphics[width=0.48\textwidth]{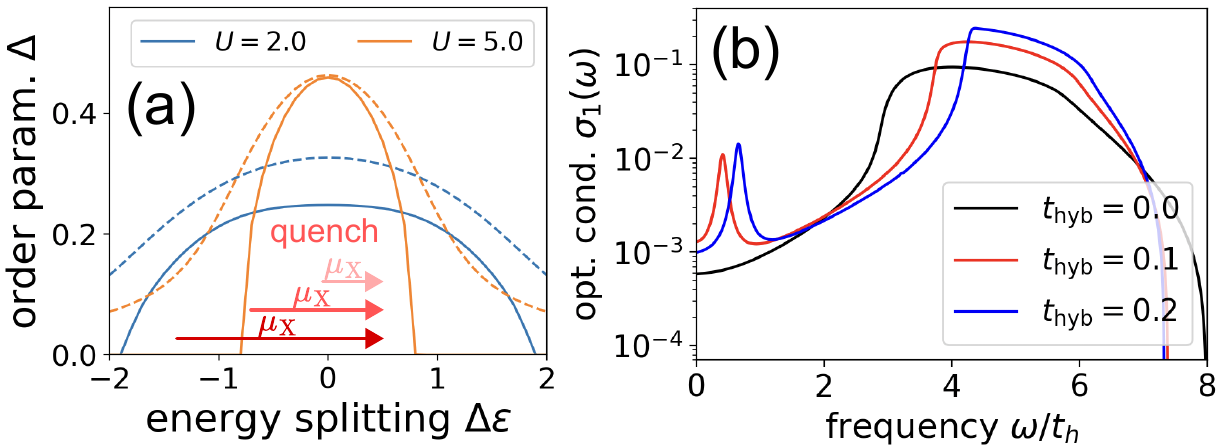}
\caption{(a) Equilibrium order parameter as a function of the orbital energy splitting $\Delta\epsilon$, solid lines have $t_\text{hyb} = 0$, dashed lines $t_\mathrm{hyb} = 0.1$. The arrow sketches the quench protocol (darker colors represent larger values of $\mu_\mathrm{X}$).
(b) Real part of the equilibrium optical conductivity $\sigma_1(\omega)$ for different values of the inter-layer tunneling $t_{\text{hyb}}$ and $U = 5$, $\Delta\epsilon = 0.6$.
The low-energy peaks appearing for $t_{\text{hyb}} > 0$ correspond to a massive phase mode.
\label{fig:opt_cond}}
\end{figure}

\paragraph{Model.}
We will restrict to a minimal model capturing the essence of a coupled bilayer by considering one band in each of the layers, which we label as $v$(alence) and $c$(onduction) band electrons.
Taking into account band inversion and a local inter-band interaction $U$ gives rise to the following Hamiltonian:
\begin{align}\begin{split}
\hat H_\text{hop} &= -t_\mathrm{h} \sum_{\langle i,j\rangle}  \big( c_i^\dagger c_j - v_i^\dagger v_j \big) \\
\hat H_{\text{loc}} &= \frac{\Delta\epsilon}{2} \sum_i\big( c_i^\dagger c_i - v_i^\dagger v_i \big) + t_\text{hyb} \sum_i\big( v_i^\dagger c_i + c_i^\dagger v_i \big) \\
&\quad + U \sum_i\Big( v_i^\dagger v_i - \frac{1}{2} \Big) \Big( c_i^\dagger c_i - \frac{1}{2} \Big).
\label{eq:hamiltonian}
\end{split}\end{align}
Here, $\langle i, j\rangle$ denotes nearest neighbours, $\Delta\epsilon$ is the energetic distance between the non-interacting bands and $t_\text{hyb}$ is the strength of the inter-layer tunnelling.
To include interaction effects in a nonequilibrium setup, we will solve the problem within tdDMFT~\cite{georges1996,aoki2014_rev} on the infinitely-connected Bethe lattice using the second-order self-consistent expansion as an impurity solver, see Ref.~\cite{SUPP} for details.
In order to be able to unambiguously resolve tunneling-induced in-gap features in the optical conductivity, we choose $U = 5$ and $\Delta \epsilon=0.6$ in the following, which gives rise to a large optical gap.
All energy~(time) scales are measured in units of (inverse) hoppings $t_\mathrm{h}~(t_\mathrm{h}^{-1})$ and the equilibrium temperature is $\beta = (20t_\text{h})^{-1}$.
We use the simplest gauge-invariant formulation of the electron-light coupling~\cite{boykin2001,golez2019a,li2020,schuler2021,dmytruk2021} leading to intraband Peierls substitution $-t_\text{h} c_i^\dagger c_j \mapsto -t_\text{h} \text{e}^{i A(t)} c_i^\dagger c_j$.
We also assume that the interband transition is dipole active, leading to an additional Hamiltonian term: 
\begin{align}\begin{split}
 \hat H_\text{dip}(t) = -D E_{\text{pr}}(t) \big( v_i^\dagger c_i + c_i^\dagger v_i \big) =: - E_{\text{pr}}(t) \hat P ,
\end{split}\end{align} where $E_{\text{pr}}=-\partial_t A$  is the  applied in-plane electric field and $\hat P$ is the corresponding in-plane polarization operator.
This coupling implies that the Hamiltonian is not centrosymmetric, which could be realized by, e.g., $s$-orbitals in one layer and $p$-orbitals in the other whose centers are not aligned.
We take $D = 0.1$ in this work.
The total induced current $j=j_{\text{intra}}+j_{\text{inter}}$ consists of an inter-band component $j_{\text{inter}}=\partial_t \langle \hat P\rangle$ and the intra-band contribution coming from the Peierls phase in the kinetic term of Eq.~(\ref{eq:hamiltonian})~\cite{aoki2014_rev,Werner2019,SUPP}
The explicit dipolar coupling $D$ is necessary for phase mode signatures to be visible in the optical spectrum, and both the bare contribution and contribution from the intralayer excitons would only lead to the renormalization of the dipolar strength~\cite{yu2015,ruiz-Tijerina2019}. 
We will evaluate the optical response in- and out-of-equilibrium directly from the response current generated by a short electric field pulse~\cite{shao2016,eckstein2008,lenarcic2014} as
$\sigma(t_{\text{pr}},\omega)=\text{FT}[j](t_\text{pr},\omega)/\text{FT}[E_\text{pr}](t_\text{pr},\omega)$, where $j(t)$ is the current responding to the field $E_\text{pr}(t) = E_\text{pr}^0 \text{e}^{-(t-t_\text{pr})^2/(2\sigma_\text{pr}^2)}$. 
Here, we have defined the Fourier transform as $\text{FT}[X](t_\text{pr},\omega)=\int_0^{t_\mathrm{cut}} X(t_{\text{pr}}+s)e^{-(i\omega+\eta) s} \text{d}s$.
The resulting data depends only weakly on the probe pulse width $\sigma_\text{pr}$, so we always choose $\sigma_\text{pr} = 0.2$.
Furthermore, we use $E_\text{pr}^0 = 10^{-2}$ and $t_\text{cut} = 80$ as a typical upper time limit for the Fourier transform and $\eta$ is a broadening factor.

\paragraph{Equilibrium and voltage-biased states.}
The Hamiltonian \eqref{eq:hamiltonian} with $t_{\text{hyb}}=0$ separately conserves the number of electrons in each band, implying invariance under changes in the $U(1)$ phase conjugate to the difference in population between the two bands. At low $T$, the symmetry may be spontaneously broken via a transition into an excitonic insulator state with the order parameter $\Delta = \langle v_i^\dagger c_i \rangle=|\Delta| e^{i \theta}$,
where the energy is independent of $\theta$ giving rise to gapless fluctuations.
A $t_\text{hyb}\neq0$ explicitly breaks this internal $U(1)$ symmetry, generically implying that $|\Delta|\neq 0$  and the energy depends on the value of $\theta$ resulting in a gapped phase mode~\cite{guseinov1973,littlewood1996,Golez2020,Murakami2020}.
In Fig.~\ref{fig:opt_cond}(a), we show the equilibrium value of $|\Delta|$ as a function of the band splitting $\Delta\epsilon$ for $t_\text{hyb}=0$ and $t_\text{hyb}\neq 0$.
In the former case, there is a critical value of band splitting $|\Delta \epsilon|$ beyond which $|\Delta|= 0$;
if $t_\text{hyb}\neq 0$, $|\Delta|\neq 0$ for all $|\Delta \epsilon|$.
Fig.~\ref{fig:opt_cond}(b) shows the equilibrium linear response optical conductivity for $U = 5$ and $\Delta\epsilon = 0.3$.
At $t_\text{hyb}=0$, the main feature is an interband transition for $\omega\approx 3 t_\text{h}$ which is greater than $2|\Delta|$ as we are not in the BCS regime.
The gap increases as $t_\text{hyb}$ increases, and an in-gap feature arising from the gapped phase mode appears.
In this situation, the strength of the in-gap feature is set by the dipolar matrix element $D$.

Next, we turn to the bilayer system in the presence of either a bias voltage or photo-excited states with quasi-stationary carrier distribution~\cite{Keldysh1986,Haug2009,reutzel2024,murotani2019,pareek2024}.
In both cases, the state is described by an inter-orbital chemical potential difference $\mu_\text{X}$.
We follow the protocol where the equilibrium state is prepared with $\Delta \epsilon=\Delta \epsilon_0-\mu_\text{X}$ and then suddenly changed to $\Delta\epsilon=\Delta\epsilon_0$ at time $t=0$ as marked in Fig.~\ref{fig:opt_cond}(a).
It was shown in Ref.~\cite{Perfetto2019} that such protocol leads to efficient real-time preparation of states with biased chemical potentials.
In the absence of inter-layer tunnelling, $t_{\text{hyb}}=0$, the number of excited carriers $N_\mathrm{X}=\sum_i \big( c_i^\dagger c_i - v_i^\dagger v_i \big)$ is a conserved quantity for all times.
If $0 < t_\mathrm{hyb} \ll t_\mathrm{h}$, $N_\text{X}$ oscillates coherently around a non-zero mean and is conserved only on average.
We confirmed numerically that these oscillations are undamped for $U = 5$ such that they give rise to a long-lived prethermalized state.
In particular, the correction to the Hartree-Fock self-energy from the second Born approximation is small in this regime due to the large gap size, which also justifies the time-dependent mean-field approximation used in previous studies~\cite{Perfetto2019,Murakami2017,Murakami2020}. Further reduction of the interband interaction $U$ leads to a decay of the oscillations of $N_\mathrm{X}$ and to a break-down of the biased state~\cite{SUPP}.

\begin{figure}
\centering
\includegraphics[width=0.48\textwidth]{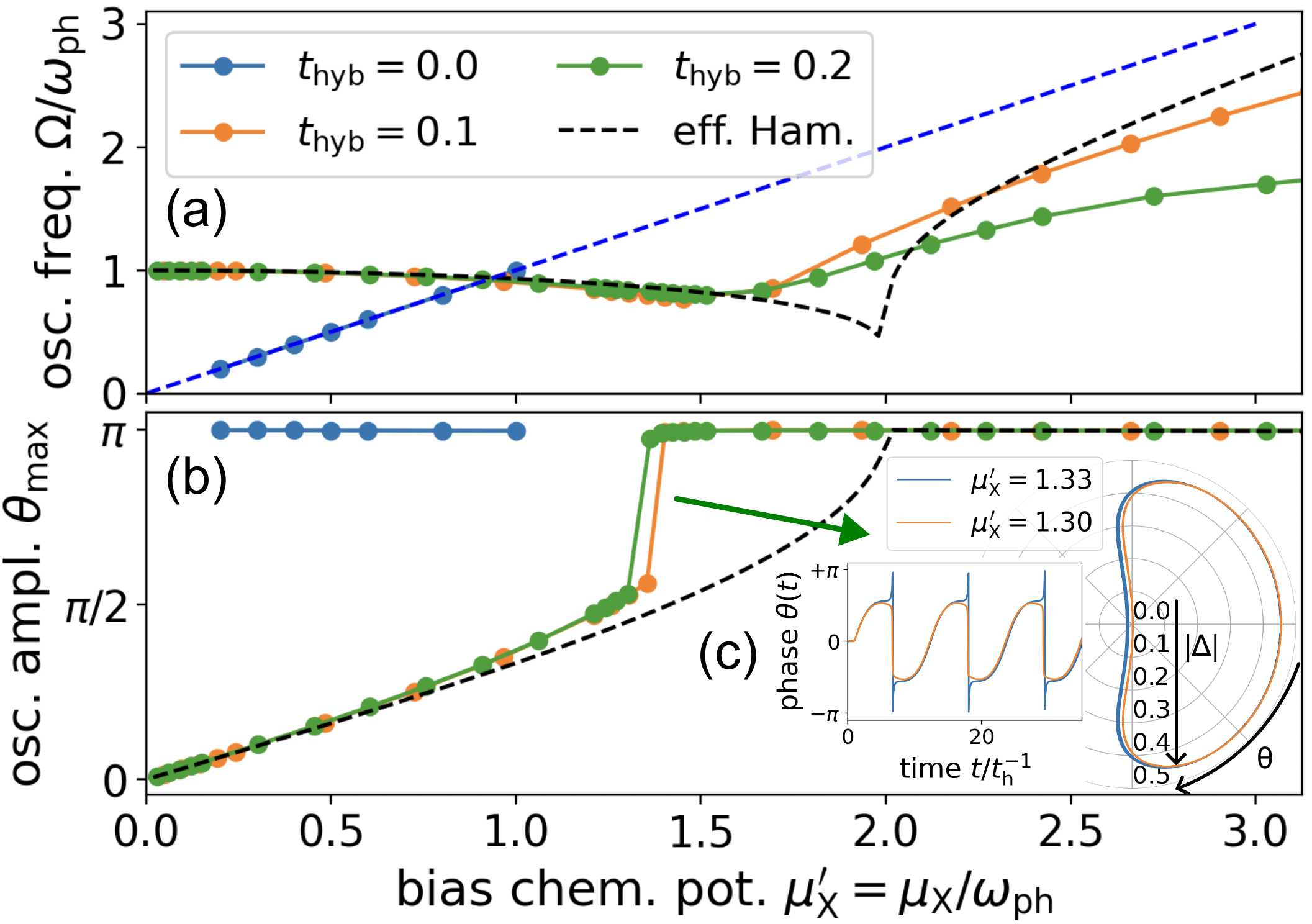}
\caption{ (a) Phase mode frequency~(confined regime) or velocity~(deconfined regime) $\Omega$ and (b) maximal amplitude $\theta_\text{max} = | \max_t \theta(t) |$ of the phase mode $\theta(t)$ versus the exciton chemical potential $\mu_\mathrm{X}$ for $U = 5$, $\Delta\epsilon = 0.6$ and varying values of $t_\mathrm{hyb}$.
The dashed lines are computed from $H_{\text{eff}}$ in~\eqref{eq:H_eff}.
For data with $t_\text{hyb} > 0$, $\mu_\text{X}$ and $\Omega$ are rescaled by the respective equilibrium phase mode frequency $\omega_\text{ph}(t_\text{hyb})$.
(c) Real-time and phase space trajectories for two values of 
$\mu_\mathrm{X}/\omega_\text{ph} = 1.30, 1.33$ ($t_\mathrm{hyb} = 0.2$)
slightly below and above the trapping transition, respectively.}
\label{fig:effham_vs_dmft}
\end{figure}

\paragraph{Order parameter dynamics.}
We begin our discussion of the post-quench dynamics with the order parameter.
Expectations about the behavior of its phase $\theta(t)$ may be formulated based on a simple time-dependent Landau-Ginzburg theory~\cite{Sun2023,Sun2021,Golez2020} if a frozen amplitude of the excitonic order parameter, $\partial_t |\Delta| = 0$, is assumed.
Then, the effective Hamiltonian for $\theta(t)$ and for our quench setup is given by~\cite{SUPP}
\begin{align}\begin{split}
 H_\text{eff} = \frac{1}{2} \dot\theta^2 + \omega_\text{ph}^2 \cos(\theta), \quad
 \begin{cases} \theta(t=0)=0 & \\ \dot\theta(t=0)=\mu_\text{X} & \end{cases} ,
 \label{eq:H_eff}
\end{split}\end{align}
where the frequency of the phase mode $\omega_{\text{ph}}(t_\text{hyb})$ is a function of the inter-layer tunnelling $t_\text{hyb}$ with $\omega_\text{ph}(t_\text{hyb} = 0) = 0$.
The voltage bias $\mu_\text{X}$ enters via the initial condition $\dot \theta(t=0)=\mu_X$, i.e. it provides an initial kick to the phase mode. The response to the probe pulse is introduced by adding a dipolar excitation term $H_\text{eff,dip}=\gamma E_{\text{pr}}(t) \cos(\theta)$, where $\gamma$ is proportional to the dipolar matrix element $D$.
If $\omega_\text{ph} = 0$, $\theta$ performs a free rotation of the form $\theta(t) = \mu_\text{X} t$, which is a direct consequence of the $U(1)$ phase translation symmetry present in this case (massless Goldstone mode).
However, for $\omega_\text{ph} > 0$,
the phase mode moves in a cosinusoidal potential and may remain trapped in the minimum around its initial value $\theta(0) = 0$ if its energy is too small.
We may therefore anticipate a transition from -- in general non-harmonic -- oscillations with $\theta_\mathrm{max} := |\max{\theta(t)}| < \pi$ at small $\mu_\text{X}$ to a winding phase motion at large $\mu_\text{X}$.
In Fig.~\ref{fig:effham_vs_dmft}(a), we plot the frequency $\Omega$ and in Fig.~\ref{fig:effham_vs_dmft}(b) the maximum value $\theta_\mathrm{max}$ of the phase oscillation.
The dashed lines corresponding to the minimal model indeed show that for $\mu_\text{X} < \mu_\text{X,c} = 2\omega_\text{ph}$, the phase mode is trapped.
The frequency $\Omega = \frac{2\pi}{T}$ of the oscillations softens upon approaching $\mu_\text{X,c}$ because of the non-harmonic parts of the cosinusoidal potential.
For $\mu_\text{X} > 2\omega_\text{ph}$, $\theta(t)$ is winding and $\Omega \sim \mu_\text{X}$ in the regime $\mu_\text{X} \gg 2\omega_\text{ph}$.

Turning to the tdDMFT simulations, represented by the coloured dots in Fig.~\ref{fig:effham_vs_dmft}, we confirm the phase rotation with frequency $\mu_\mathrm{X}$ for $t_\text{hyb} = 0$.
If $t_{\text{hyb}} > 0$, the predicted softening of $\Omega$ at small values of $\mu_\text{X}$ is present as well.
However, the untrapping transition occurs sharply at a chemical potential smaller than $2\omega_\text{ph}$.
To shed more light on this behaviour, we plotted $\theta(t)$ and the $|\Delta|$-$\theta$ phase space trajectory in Fig.~\ref{fig:effham_vs_dmft}(c) for values of $\mu_\mathrm{X}$ slightly below and slightly above the transition.
In the phase space picture, the shape of the trajectory is strongly squeezed along the $|\Delta|$-direction close to $|\Delta| = 0$.
This explains the possibility of a sudden strong change of the range of explored angles.
Since the minimal model~\eqref{eq:H_eff} assumes a fixed value of the amplitude of the order parameter $|\Delta| = \text{const.}$, i.e. circle arcs in the phase space picture, as well as neglects higher-order gradient term, such dynamics cannot occur there, leading to qualitatively different response than in the microscopic evolution. However, one can roughly reproduce the squeezed trajectories as equipotential lines within the equilibrium Landau-Ginzburg potential~\cite{SUPP}.

\begin{figure}
\centering
\includegraphics[width=0.48\textwidth]{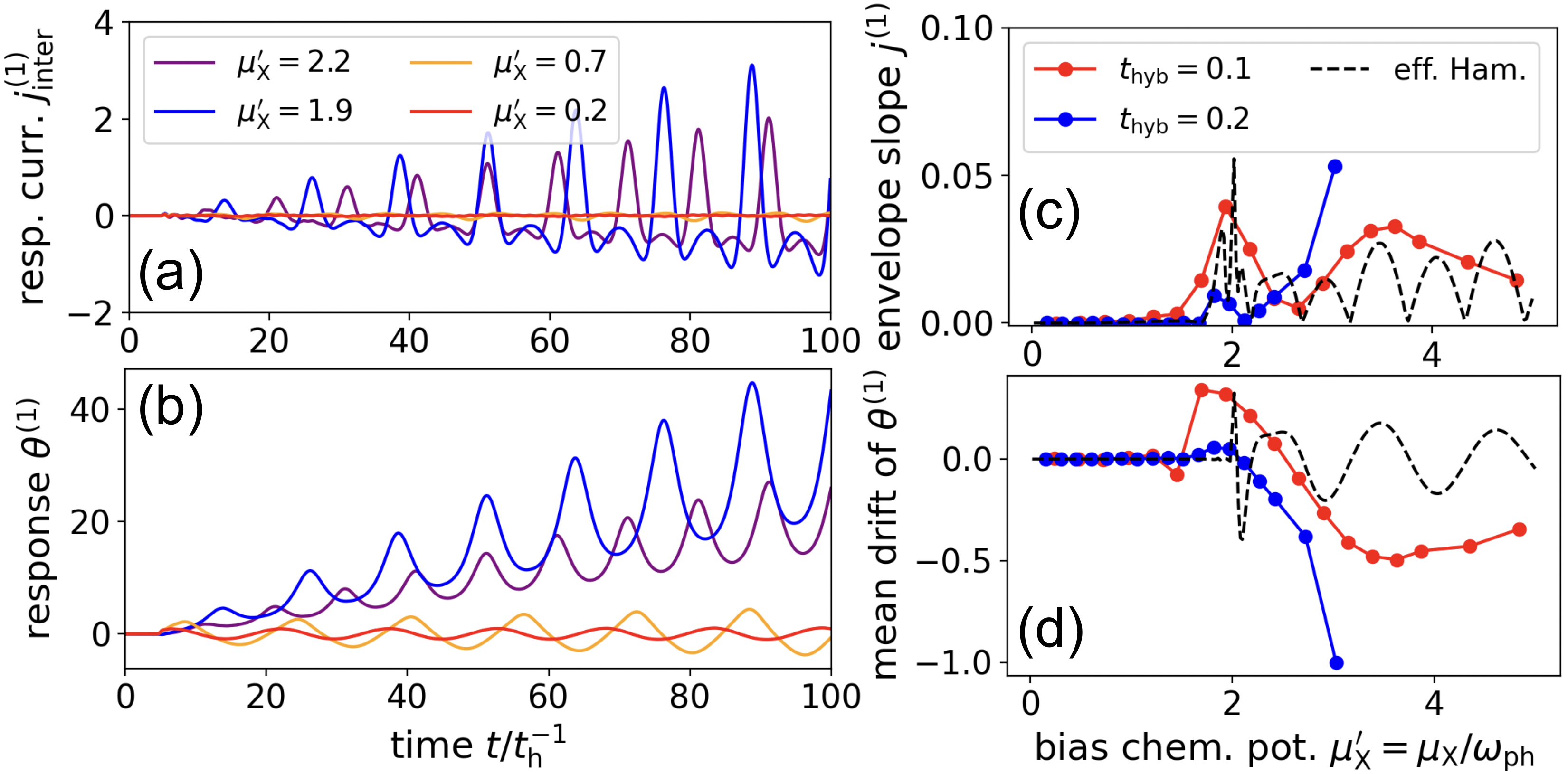}
\caption{(a) The inter-layer current response $j_\mathrm{inter}^{(1)}(t)$ and (b) the order parameter's response $\theta^{(1)}(t)$ of the dynamical condensate after the application of a weak Gaussian-shaped probe pulse.
(c) Slope of a linear fit to the maxima of $j_\mathrm{inter}^{(1)}(t)$ (envelope) and (d) slope of a linear fit to the phase response of the order parameter~$\theta^{(1)}$ versus biased chemical potential $\mu_x$. System parameters are $U = 5$, $\Delta\epsilon = 0.6$ and $t_\text{hyb} = 0.1$ and for the effective Hamiltonian model we assume $|\Delta| = 0.4$.
\label{fig:resp_currents}}
\end{figure}

\paragraph{Optical response:~currents.}
We will now argue that the two dynamical regimes can also be distinguished in optical measurements.
Using $j_\text{inter}(t) = \partial_t \langle P \rangle = 2 D \big( (\partial_t | \Delta |) \cos(\theta) - | \Delta | \dot\theta \sin(\theta) \big)$,
Fig.~\ref{fig:resp_currents}(a) shows the inter-layer response current $j_\text{inter}^{(1)}(t) = \big( j_\text{inter}(t)|_{E_{\text{pr}}^0 > 0} - j_\text{inter}(t)|_{E_{\text{pr}}^0 = 0} \big) / E_{\text{pr}}^0$, obtained from the tdDMFT simulation for values of $\mu_\mathrm{X}$ both in the trapped and the winding regime of the phase mode.
In the first case, the response current remains small, while it grows strongly with a linearly growing envelope in the latter case.
A more quantitative analysis of this observation yields Fig.~\ref{fig:resp_currents}(c), in which the slope of the envelope of $j^{(1)}(t)$ is plotted. In contrast to the jump for the phase of the order parameter $\theta$ in Fig.~\ref{fig:effham_vs_dmft}(b), the physically observable current response grows more smoothly, but with a significant increase of the slope beyond the critical $\mu_\text{X,c}$, see Fig.~\ref{fig:resp_currents}.
The growth of the inter-layer response current is connected to the
phase response of the order parameter, $\theta^{(1)}(t)$,
shown in Fig.~\ref{fig:resp_currents}(b). 
In the winding phase mode regime, we find $\theta^{(1)}(t)$ oscillating around a drifting mean while it oscillates around zero in the trapped regime. Fig.~\ref{fig:resp_currents}(d) displays the drift velocity as a function of $\mu_\text{X}$ and shows qualitative agreement with the effective model.
Hence, the probe-induced current's behaviour goes hand-in-hand with the strong growth of the order parameter's phase such that these quantities can detect the presence of the dynamical transition.

\begin{figure}
\centering
\includegraphics[width=0.48\textwidth]{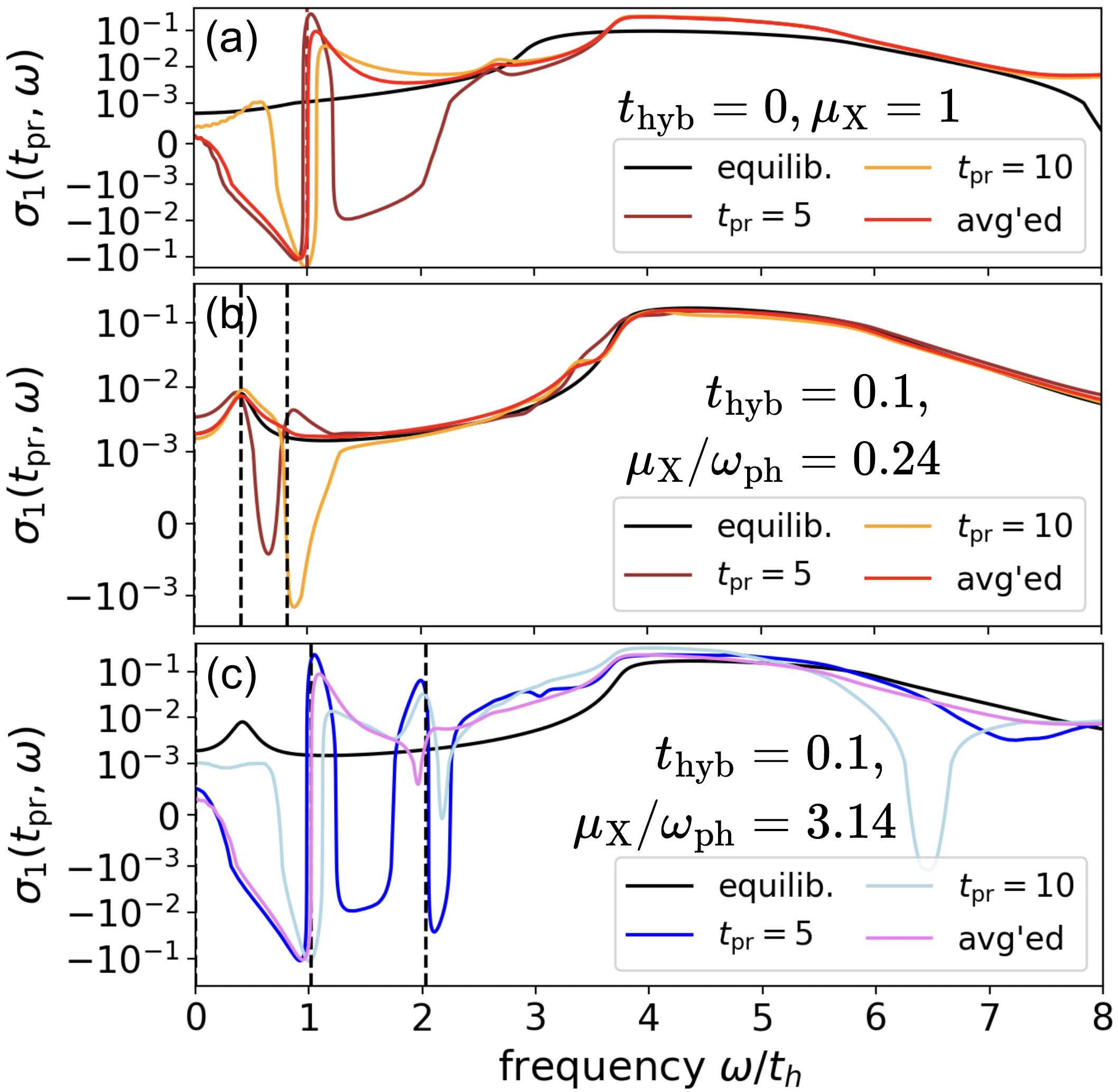}
	\caption{Non-equilibrium optical conductivity $\sigma_1(t_\mathrm{pr}, \omega)$ for (a) decoupled layers $t_\mathrm{hyb}=0$ after a quench of size $\mu_\mathrm{X}=1$ and for coupled layers with $t_\mathrm{hyb}=0.1$ in the (b) phase-trapped, $\mu_\text{X}/\omega_\text{ph} = 0.24$ ($\mu_\mathrm{X}=0.1$), and (c) phase-delocalized, $\mu_\text{X}/\omega_\text{ph} = 3.14$ ($\mu_\mathrm{X}=1.3$), regimes at different delay times $t_\text{pr}$.
Dashed lines show the main oscillation frequency of $\theta(t)$, which agrees with the equilibrium phase mode peak position only in the confined case and its second harmonics. The used broadening for the Fourier transform is $\eta  = 10 / t_\text{cut}$.
\label{fig:op_transition_curr}}
\end{figure}

\paragraph{Optical response:~conductivities.}
Finally, we discuss the Fourier-transformed total response current and the time-dependent optical conductivities $\sigma(t_{\text{pr}},\omega)$.
We begin with the case of decoupled layers $t_\mathrm{hyb} = 0$, shown in Fig.~\ref{fig:op_transition_curr}(a) exemplarily for a value of $\mu_\mathrm{X} = 1$.
One finds that the dynamical phase $\theta(t) = \mu_\mathrm{X} t$ indeed shows up in the (non-equilibrium) optical conductivity $\sigma_1(t_\mathrm{pr}, \omega)$ as an additional peak at $\omega = \mu_\text{X}$.
In particular, the position of the peak is independent of the precise probe time delay $t_\mathrm{pr}$, while the optical weight depends on it (within a period $T = \frac{2\pi}{\mu_\text{X}}$) and can become negative as observed in other contexts~\cite{Chiriaco2020,Golez2020,golez2017}. 
This can be schematically understood from the dynamics of the response current in Fig.~\ref{fig:resp_currents}(a)
since the Fourier transform of a current $j(t) \sim t \sin(\nu t)$ has components $\pm \partial_\omega\delta(\omega \mp \nu)$.
Negative optical conductivities indicate the system's tendency for stimulated emission of photons at a given frequency.
We also show the period-averaged optical conductivity $\overline{\sigma}(\omega) = \frac{1}{T}\int_{t_0}^{t_0+T} \sigma(t_\mathrm{pr}, \omega) \text{d}t_\mathrm{pr}$ as a measure of average absorption/emission in typical stroboscopic experimental setup.

Optical conductivities at finite interlayer tunnelling $t_\mathrm{hyb}=0.1$ are shown in Fig.~\ref{fig:op_transition_curr}(b) for the trapped-phase and Fig.~\ref{fig:op_transition_curr}(c) for the phase-delocalized regime, respectively.
In the trapped state with $\mu_\mathrm{X}/\omega_{\text{ph}} = 0.24$ ($\mu_\mathrm{X} = 0.1$), we find a peak appearing at the equilibrium phase mode frequency $\omega_\text{ph}$ which is different from $\mu_\text{X}$. In contrast, in the regime with complete phase winding, $\mu_\mathrm{X}/\omega_{\text{ph}} = 3.14$ ($\mu_\mathrm{X}=1.3$), $\sigma_1(t_\mathrm{pr},\omega)$ does not peak at the phase mode frequency but instead at a larger frequency corresponding to the main oscillating frequency of the order parameter's phase.
Again, these signals can have positive or negative weight, depending on the probe time.
The period-averaged optical conductivity for the trapped phase in Fig.~\ref{fig:op_transition_curr}(b) remains positive at all frequencies. In both regimes, the second harmonics at $2\omega_\text{ph}$ appear, see dashed lines in Fig.~\ref{fig:op_transition_curr}(b)-(c), in agreement with previous studies~\cite{Golez2020}. 
In-gap states in the optical response are therefore not only a signature of the \emph{dynamical excitonic condensation}, but their scaling with the strength of the excitation $\mu_{X}$ allows, in addition, to distinguish between different dynamical regimes of excitonic order.

\paragraph{Conclusions.}
We studied the optical properties of a dynamical exciton condensate as realized either in bilayer structures with an applied voltage bias or photodoped semiconductors.
The dynamical condensate may be in one of two states, characterized by the long-time behaviour of the phase of the condensate amplitude.
In the weakly nonequilibrium state, the phase is confined, oscillating about the value that minimizes the free energy.
In the strongly nonequilibrium state, the phase is deconfined, linearly increasing with time. We used nonequilibrium dynamical mean-field theory to understand the two states, the transition between them, and the observable signatures in the linear response optical conductivity.
At small voltage bias $\mu_\text{X}$, the optical conductivity shows a sharp peak at the energy of the equilibrium phase mode excitation.
At large voltage bias $\mu_\mathrm{X}$,
the low-energy optical peak decouples from the phase mode frequency and a wide regime of stimulated optical emission appears.
The evolution of the in-gap peak with increasing driving strength thus provides experimental access to detect the state of dynamical condensate.

An important direction for future research concerns the nature of spontaneous emission in driven condensates,
which would require to solve the coupled evolution of the electromagnetic environment and the correlated solid.
A further extension is to modify the electromagnetic environment by placing the material in a cavity in order to understand if the super-radiant phase proposed in effective theories~\cite{Sun2023,littlewood1996} is consistent with the microscopic evolution presented in this work.
Finally, the nonlinear response, such as the order of the Josephson's effect~\cite{Sun2021} and nonlinear optical response~\cite{kaneko2021},  sensitively depends on the microscopic details, and it would be important to extend the current microscopic description to realistic setups.

\begin{acknowledgments}
{\it Acknowledgements.}
We acknowledge discussions with G. Mazza and M. Michael. A.O. and D.G. acknowledge support from No. P1-0044, No. J1-2455, No. J1-2458 and No. MN-0016-106 of the Slovenian Research Agency (ARIS) and QuantERA grants QuSiED by MVZI and QuantERA II JTC 2021.
Y.M. and T.K. are supported by Grant-in-Aid for Scientific Research from JSPS, KAKENHI Grant No.~JP20H01849 (T.K.), No.~JP21H05017 (Y.M.), No.~JP24K06939 (T.K.), No.~JP24H00191, JST CREST Grant No.~JPMJCR1901 (Y.M.), and the RIKEN TRIP initiative RIKEN Quantum (Y.M.). Z.S. is supported by the National Natural Science Foundation of China (Grants No. 12374291 and No. 12421004). A.J.M. acknowledge support from Programmable Quantum Materials, an Energy Frontiers Research Center funded by the U.S. Department of Energy (DOE), Office of Science, Basic Energy Sciences (BES), under award DE-SC0019443.
The Flatiron institute is a division of the Simons foundation.
The NESSi package~\cite{NESSI} was used for the calculations in this paper.
\end{acknowledgments}

\bibliography{main}

\end{document}


\title{Supplement: Optical signatures of dynamical excitonic condensates}

\maketitle

\section{Details on the numerics}

We solve the problem within the time-dependent dynamical mean-field theory~\cite{aoki2014_rev} on the Bethe lattice using a self-consistency condition with time-dependent vector potential~\cite{Werner2017}:
\begin{align}\begin{split}
 \underline{\underline{\Delta}}(t, t') &= \frac{1}{2} \Big( \underline{\underline{v_i}}(t) \cdot \underline{\underline{G}}(t, t') \cdot \underline{\underline{v_i^\dagger}}(t') + \underline{\underline{v_i^\dagger}}(t) \cdot \underline{\underline{G}}(t, t') \cdot \underline{\underline{v_i}}(t') \Big) \\
 &=: \frac{1}{2} \Big( \underline{\underline{\Delta_L}}(t, t') + \underline{\underline{\Delta_R}}(t, t') \Big)
\end{split}\end{align}
with diagonal matrix (no time-dependence if the field is off)
\begin{equation}
 \underline{\underline{v_i}}(t) = \begin{pmatrix} - t_\text{h} \text{e}^{i A(t)} & 0 \\
 0 & + t_\text{h} \text{e}^{i A(t)} \end{pmatrix}.
\end{equation}

The intra-layer/band current in the Bethe lattice DMFT framework is calculated as $j_\text{intra}(t) = \operatorname{Im} \operatorname{tr} \big[ \Gamma_\text{L}(t) - \Gamma_\text{R}(t) \big]$,
using the intermediate objects $\Gamma_{L/R}(t) = -i ~ ( G \ast \Delta_{L/R} )(t)$.

First order self-energies (note $G_{f f}(t,t^+) = G_{f f}^<(t,t) = i n_f(t)$) for $f = c, v$
\begin{align}\begin{split}
 \Sigma_{f,f}^\text{H}(t, t') &= -i U(t) \delta_C(t, t') G_{\bar f,\bar f}(t,t^+) = + U(t) \delta_C(t, t') n_{\bar f}(t) \hspace{1cm} \text{(Hartree)}, \\
 \Sigma_{f,\bar f}^\text{F}(t, t') &= + i U(t) \delta_C(t, t') G_{f,\bar f}(t, t^+) = - U(t) \delta_C(t, t') \langle {\bar f}^\dagger f \rangle(t) \hspace{0.7cm} \text{(Fock)} ,
\end{split}\end{align}
where $\bar f$ marks the opposite band.
Next to these Hartree and Fock diagrams, we will include the two second-order skeleton self-energy diagrams (2nd Born approximation):
\begin{align}\begin{split}
 \Sigma_{f,f'}^\text{2nd,1}(t, t') = - i^2 U(t) U(t') G_{f,f'}(t, t') G_{\bar f, \bar f'}(t,t') G_{\bar f',\bar f}(t', t) , \\
 \Sigma_{f,f'}^\text{2nd,2}(t, t') = + i^2 U(t) U(t') G_{f,\bar f'}(t, t') G_{\bar f',\bar f}(t',t) G_{\bar f,f'}(t, t') .
\end{split}\end{align}

\section{Equilibrium data}
\begin{figure}[ht]
\centering
\includegraphics[width=0.6\textwidth]{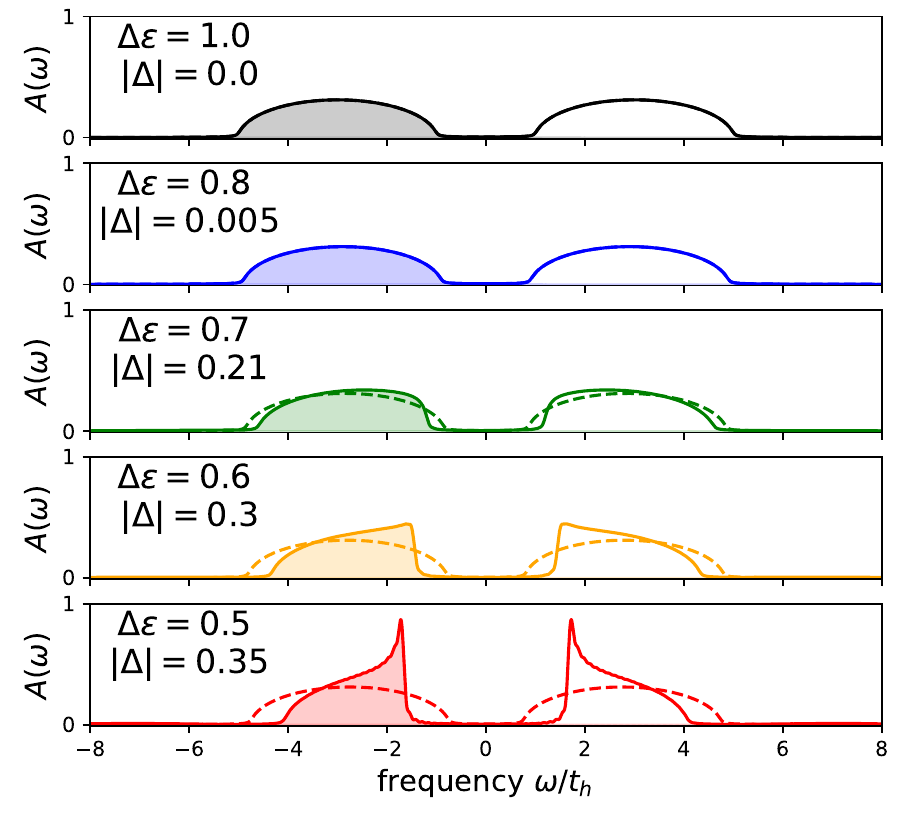}
\caption{Equilibrium spectral functions for varying values of $\Delta\epsilon$, $U/t_\text{h} = 5$ and $t_\text{hyb} = 0$.
Dashed lines show the non-hybridized orbital spectral densities for comparison\label{suppfig:specfunc-eq_comp-eps0C}.}
\end{figure}

\begin{figure}[ht]
\centering
\includegraphics[width=0.7\textwidth]{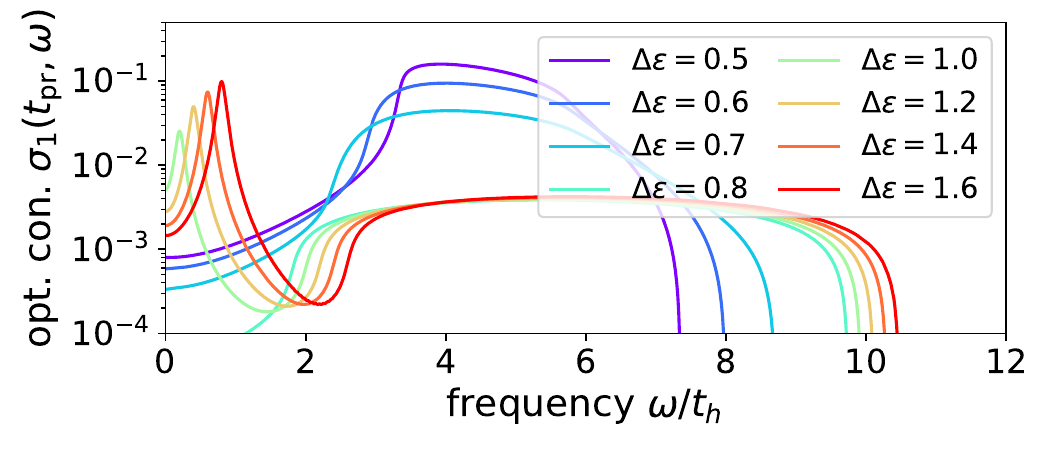}
\includegraphics[width=0.7\textwidth]{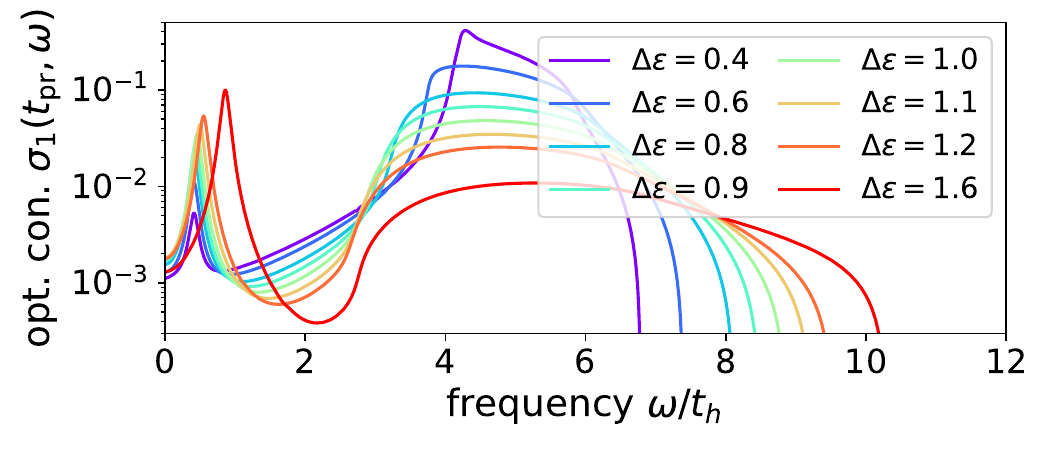}
\caption{Equilibrium optical conductivities $\sigma_1(\omega)$ for varying values of $\Delta\epsilon$, $U/t_\text{h} = 5$ and $t_\text{hyb} = 0$ (upper), $t_\text{hyb} = 0.1$ (lower)\label{suppfig:optcond-eq_2B_Uinter5}}
\end{figure}

Figure~\ref{suppfig:specfunc-eq_comp-eps0C} shows the spectral functions $A^\text{ret}(\omega)$ (density of states, displayed by the solid lines) and $A^<(\omega)$ (occupied density of states, displayed by the shaded area) for the transition from a normal state (with the semi-circular density of states of the Bethe lattice) at a large value of $\Delta\epsilon$ to the excitonic-insulating state with spectral weight accumulating at the gap edges.
The spectral functions are calculated as orbital traces from the Fourier-transformed retarded and lesser Greens' functions.
In the equilibrium case, the real-time functions only depend on $\tau = t - t'$, we provide the more general definition below: 
\begin{align}\begin{split}
 G_f^<(t, t') &= + i \big\langle c_f^\dagger(t') c_f(t) \big\rangle, \quad
 G_f^\text{ret}(t,t') = -i \theta(t-t') \big\langle \big\{ c_f(t), c_f^\dagger(t') \big\} \big\rangle, \\
 A^\text{X}(t', \omega) &= -\operatorname{Im} \sum_f \frac{1}{\sqrt{2\pi}} \int \text{d}\tau \; \text{e}^{(i \omega - \eta) \tau} G_f^\text{X}(t'+\tau,t') , \quad X \in \{ \text{ret}, < \}
\end{split}\end{align}
As $\Delta\epsilon$ decreases, the spectra deviate more and more from the non-hybridized orbitals and spectral weight accumulates more at the gap edges.
Figure~\ref{suppfig:optcond-eq_2B_Uinter5} presents the respective equilibrium optical conductivities (colors do not match) for the $U/t_\text{h} = 5$ system.
If $t_\text{hyb} = 0$ (upper panel),
there is a sharp exciton peak at low energies (although $\Delta = 0$ in this regime).
It stems from the inter-band current due to the dipolar coupling and shows the instability of the system towards exciton formation.
This peak moves to lower and lower energies as $\Delta\epsilon$ decreases until the system starts to order spontaneously ($|\Delta| > 0$) once it hits zero (cf. Fig.~\ref{suppfig:specfunc-eq_comp-eps0C}).
The formation of order is then accompanied by a significant increase of the spectral weight of the continuum.
The picture is different for $t_\text{hyb}/t_\text{h} = 0.1$ (lower panel),
since the equilibrium state has a non-zero order parameter for all $\Delta\epsilon$.
At small values of $\Delta\epsilon$, the excitonic phase mode peak at low energies does not scale to zero but remains non-zero.

\clearpage
\section{Probe-time dependence of optical conductivities}

We argued in the main text that the distribution of positive and negative weights of the non-equilibrium optical conductivity $\sigma_1(t_\mathrm{pr}, \omega)$ depends strongly on the choice of the probe time $t_\mathrm{pr}$.
In the regime, which we are studying (long-lived mean-field-like states), there is a natural periodicity in observables like the order parameter (e.g. $T = \frac{2\pi}{\mu_\text{X}}$ in the case $t_\text{hyb} = 0$),
and we confirmed that also $\sigma_1(t_\mathrm{pr}+T,\omega) = \sigma_1(t_\mathrm{pr}, \omega)$.

Therefore, it appears natural to calculate a period-averaged optical conductivity $\overline{\sigma}(t_\mathrm{pr}, \omega) = \frac{1}{T} \int_0^T \sigma(t_\mathrm{pr}+\tau, \omega) \text{d}\tau$.
One may motivate this quantity firstly via the finite coherence length in a condensate, which is smaller than the wave length of the light.
Different coherent “patches" develop a random phase relation to each other, which translates to random probe time shifts (because the phase mode frequency corresponds to the slowest timescale) over which the optical measurement averages.
As a second motivation, one may also think about repeated measurements in experimental setups (“repetition rate") that average over a random phase difference between repetitions.

\begin{figure}[ht]
\centering
\includegraphics[width=0.5\textwidth]{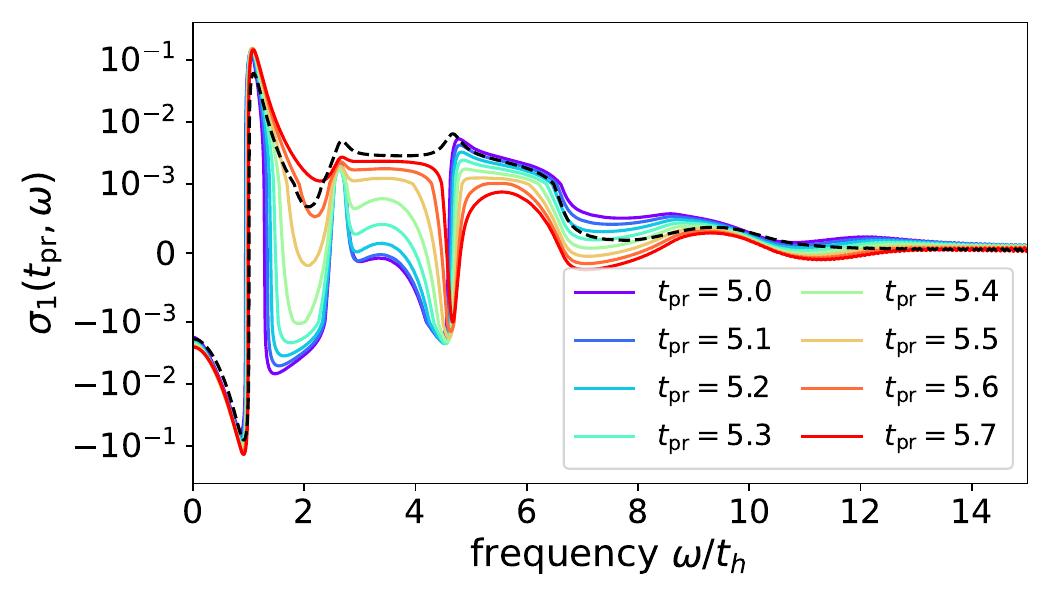}%
\includegraphics[width=0.5\textwidth]{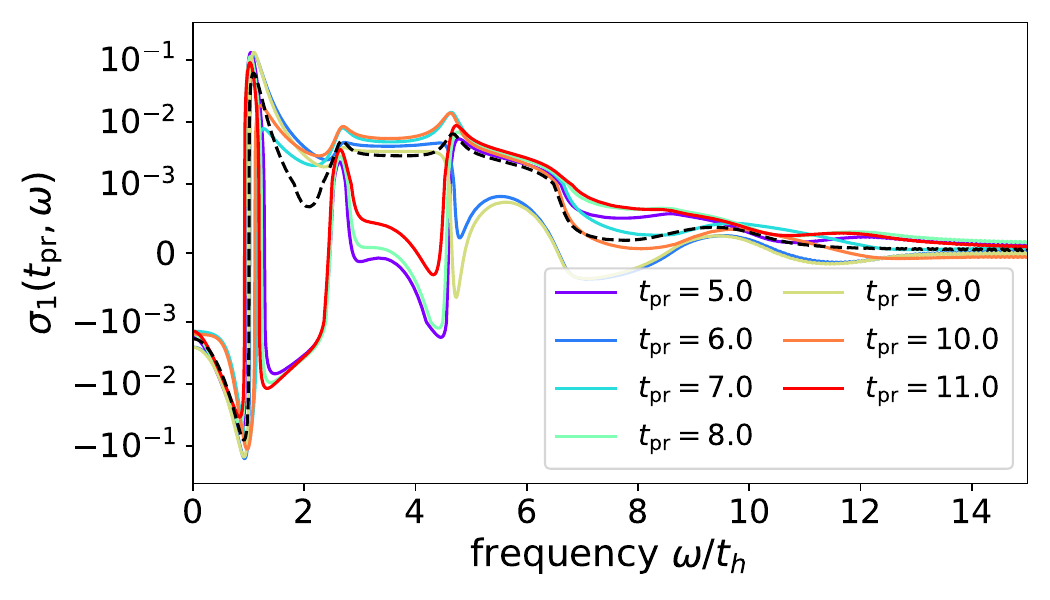}
\caption{Probe-time dependence of the optical conductivities calculated only from the inter-layer contribution to the response current, $j_\text{inter}(t)$, after a quench with $\mu_\mathrm{X} = 1.0$ and for $U = 5 t_\mathrm{h}$, $\Delta\epsilon = 0.6 t_\mathrm{h}$ and $t_\mathrm{hyb} = 0$.
The dashed line shows the period-averaged optical conductivity $\overline{\sigma}(\omega)$.
\label{suppfig:optcond_switchPM3e-1_hyb0_eps0C-2.00-01_tpr_avgd}}
\end{figure}

\begin{figure}[ht]
\centering
\includegraphics[width=0.5\textwidth]{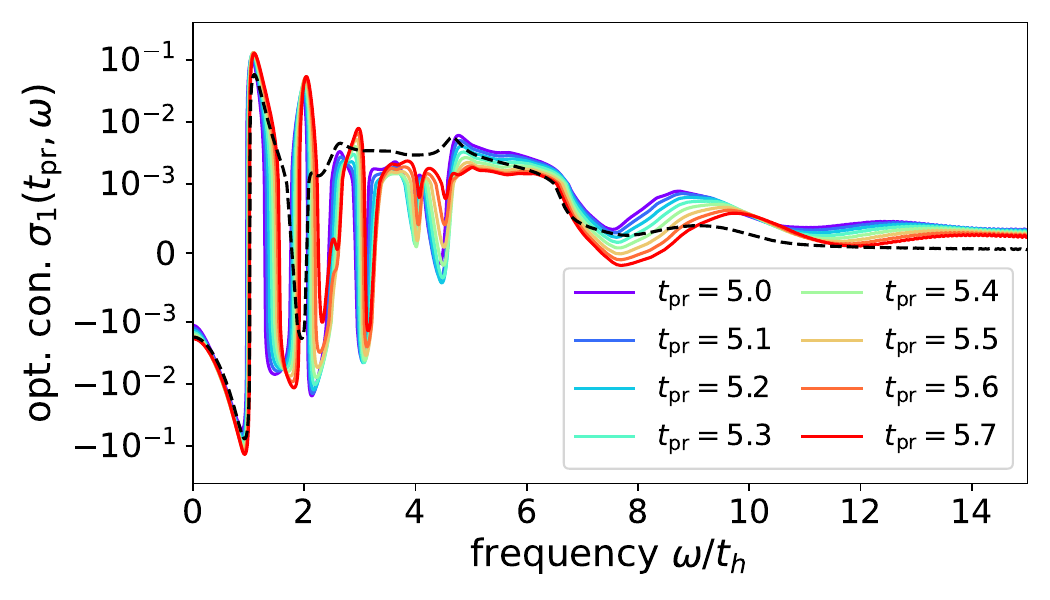}%
\includegraphics[width=0.5\textwidth]{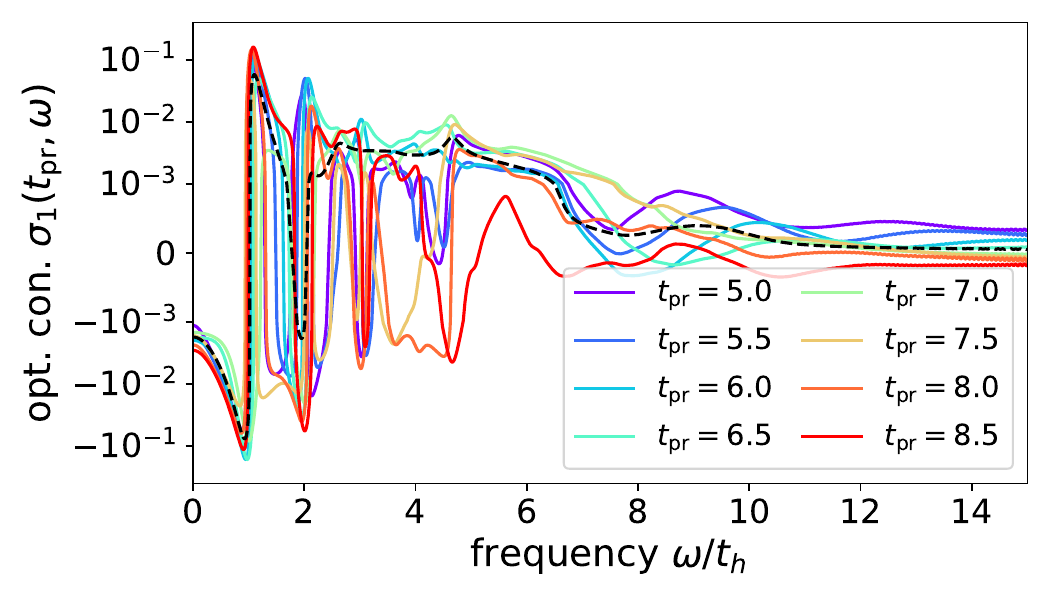}
\caption{Optical conductivities after a quench with $\mu_\mathrm{X} = 1.3$ and for $U = 5 t_\mathrm{h}$, $\Delta\epsilon = 0.6 t_\mathrm{h}$ and $t_\mathrm{hyb} = 0.1$.
\label{suppfig:optcond_switchPM3e-1_hyb1e-1_eps0C-3.50-01_tpr_avgd}}
\end{figure}

\begin{figure}[ht]
\centering
\includegraphics[width=0.5\textwidth]{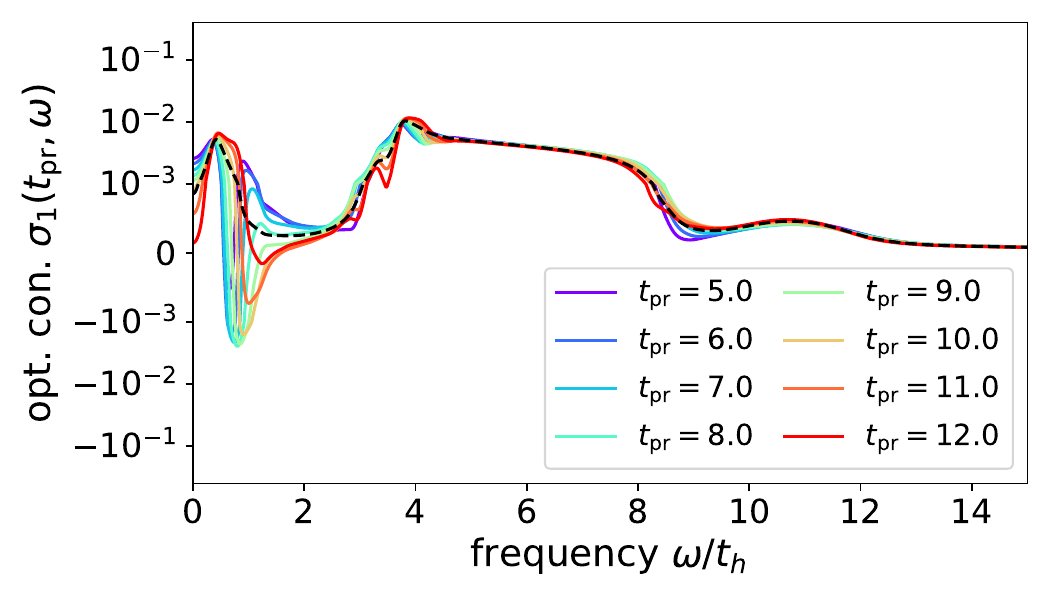}
\caption{Optical conductivities after a quench with $\mu_\mathrm{X} = 0.1$ and for $U = 5 t_\mathrm{h}$, $\Delta\epsilon = 0.6 t_\mathrm{h}$ and $t_\mathrm{hyb} = 0.1$.
\label{suppfig:optcond_switchPM3e-1_hyb1e-1_eps0C2.50-01_tpr_avgd}}
\end{figure}

Below, we show only the contribution of the optical conductivity, which stems from the inter-layer current $j_\text{inter}(t) = \partial_t \langle \hat P \rangle$, of the data shown in the main text, with more probe time data and the averaged conductivity as a dashed line.
Figure~\ref{suppfig:optcond_switchPM3e-1_hyb0_eps0C-2.00-01_tpr_avgd} is for $t_\text{hyb} = 0$.
Figures~\ref{suppfig:optcond_switchPM3e-1_hyb1e-1_eps0C-3.50-01_tpr_avgd}~and~\ref{suppfig:optcond_switchPM3e-1_hyb1e-1_eps0C2.50-01_tpr_avgd} have $t_\text{hyb} = 0.1$ with $\mu_\text{X}/\omega_{\text{ph}} = 0.24$ ($\mu_\text{X} = 0.1$) and $\mu_\text{X}/\omega_{\text{ph}} = 3.14$ ($\mu_\text{X} = 1.3$), respectively.

Numerically, these have been obtained by calculating the average over a set of $\sigma(t, \omega)$,
where $t \in [t_\mathrm{pr}, t_\mathrm{pr}+T)$ is evaluated on an evenly-spaced time grid with $\Delta t = 0.1 t_\mathrm{h}^{-1}$.

\clearpage
\section{Other values of the inter-band interaction $U$}
\begin{figure}[ht]
\centering
\includegraphics[width=0.5\textwidth]{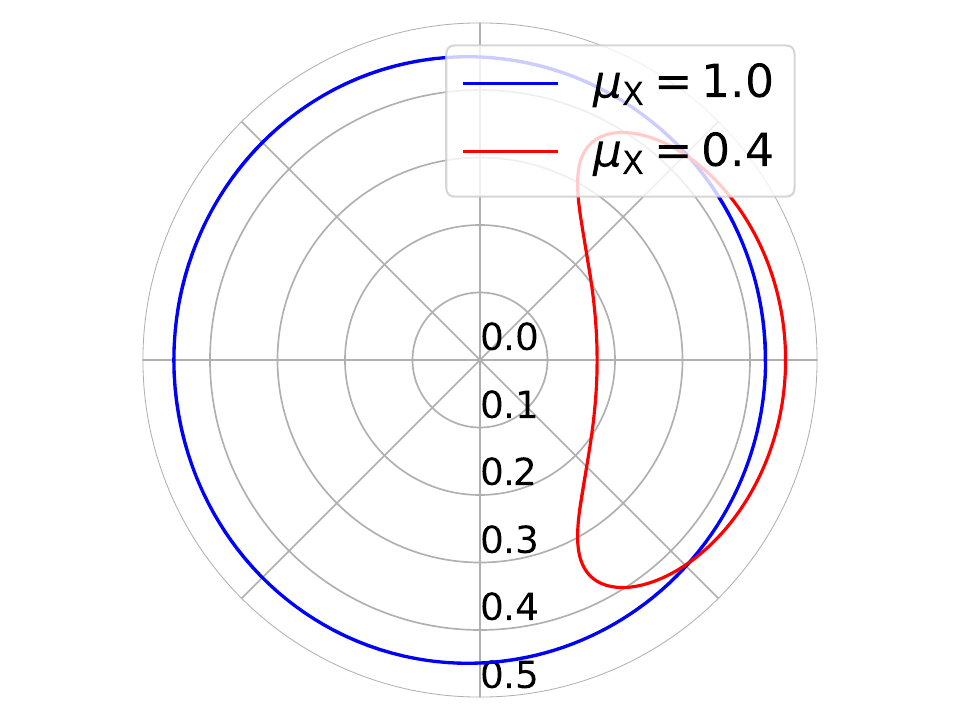}%
\includegraphics[width=0.5\textwidth]{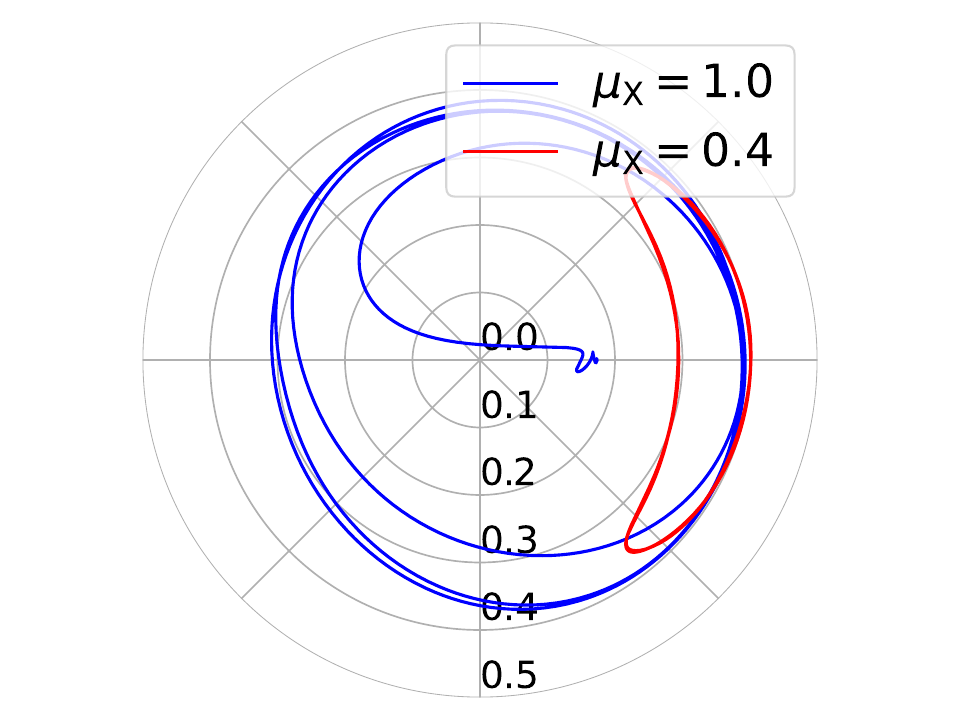}
\caption{Order parameter trajectories in phase space for $\Delta\epsilon = 0.6 t_\text{h}$, $t_\text{hyb} = 0.1 t_\text{h}$ with $U = 5 t_\text{h}$ (left) and $U = 3 t_\text{h}$ (right).
Comparison of $\mu_\text{X} = 0.4$ ($\mu_\text{X}' = 0.97$ if $U = 5$) in the trapped and $\mu_\text{X} = 1.0$ ($\mu_\text{X}' = 2.42$ if $U = 5$) in the winding regime.
\label{suppfig:op_polar_switchEpsToPM3e-1_M1W5e-2_2B_hyb1e-1_U}}
\end{figure}

In this section, we would like to contrast the behavior at $U = 5 t_\text{h}$,
on which we focused in the main text, with a smaller values of $U$.
In Fig.~\ref{suppfig:op_polar_switchEpsToPM3e-1_M1W5e-2_2B_hyb1e-1_U},
we show a direct comparison of the order parameter phase-space trajectories at $U = 5 t_\text{h}$ and $U = 3 t_\text{h}$.
For $U = 5 t_\text{h}$, we obtain the mean-field-like closed orbits, which we discussed in the main text.

\begin{figure}[ht]
\centering
\includegraphics[width=0.5\textwidth]{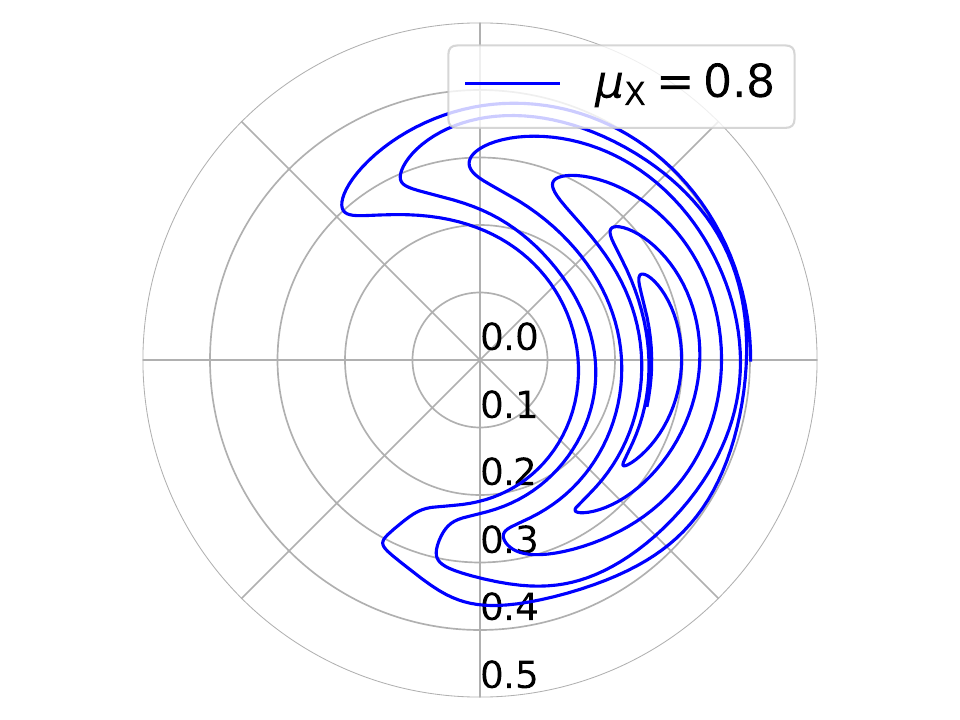}%
\includegraphics[width=0.5\textwidth]{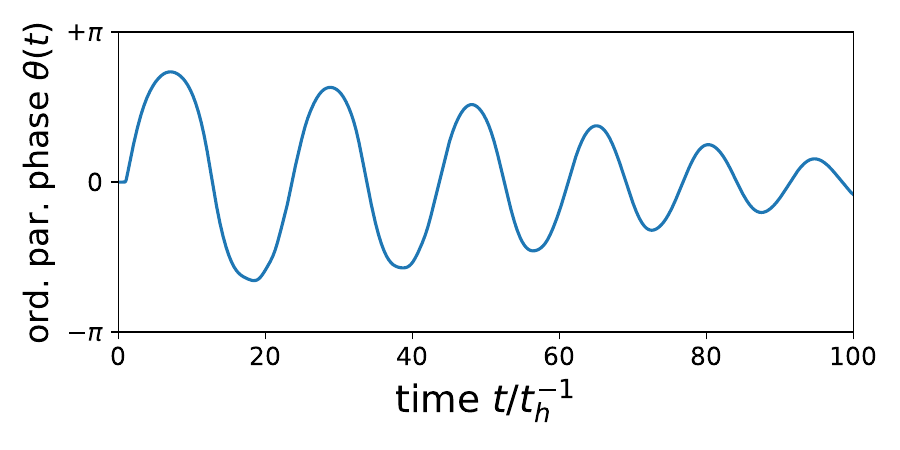}
\caption{Order parameter trajectory in phase space (left) and order parameter phase in time (right) for $\Delta\epsilon = 0.6 t_\text{h}$, $t_\text{hyb} = 0.1 t_\text{h}$ and $U = 3 t_\text{h}$.
An applied $\mu_\text{X} = 0.8$ generates a state, which is initially in the trapped regime but then decays in time.
\label{suppfig:op_polar_switchEpsToPM3e-1_M1W5e-2_2B_hyb1e-1_U3}}
\end{figure}

For $U = 3 t_\text{h}$, we find a similar closed orbit for $\mu_\text{X} = 0.4$ (trapped regime) but for the larger quench $\mu_\text{X} = 1.0$ (winding regime in this case) the phase space trajectory collapses (within the first $20 t_\text{h}^{-1}$ time units after the quench) to a static state with a decayed phase mode.
Such a decay can also occur in the trapped regime,
as we demonstrate in Fig.~\ref{suppfig:op_polar_switchEpsToPM3e-1_M1W5e-2_2B_hyb1e-1_U3} with $\mu_\text{X} = 0.8$.

\clearpage
\section{Effective Hamiltonian / Time-dependent Ginzburg-Landau theory}

\subsection{Phase space trajectories}
\begin{figure}[ht]
\centering
\includegraphics[width=0.8\textwidth]{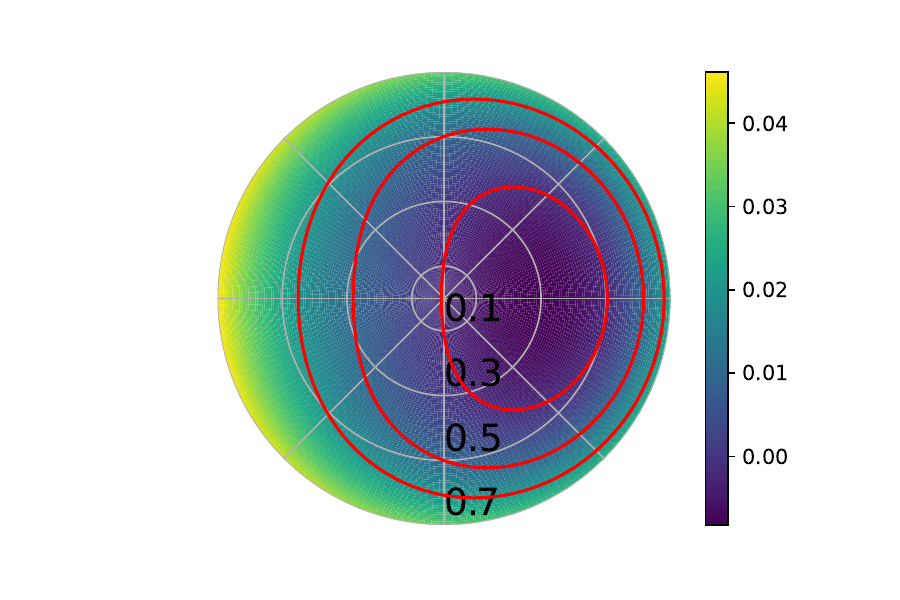}
\caption{Effective potential $V_\text{eff}\big( |\Delta|, \theta \big)$ for $U = 5$, $t_\text{hyb} = 0.1$, $a = 0.18$, $b = 0.1$ with equipotential lines at $V = -0.0015, 0.01, 0.02$.\label{suppfig:traj_gl}}
\end{figure}

If we assume a Ginzburg-Landau Lagrangian of the type
\begin{align}\begin{split}
L = \overline{\Delta} \partial_t^2 \Delta -\frac{1}{U} \big| \Delta \big|^2 + a \big| \Delta + t_\text{hyb} \big|^2 - b \big| \Delta + t_\text{hyb} \big|^4 ,
\end{split}\end{align}
one can ask for the shape of equipotential lines in the effective potential (neglecting the kinetic term for now)
\begin{align}\begin{split}
V_\text{eff}\big( |\Delta|, \theta \big) = \frac{1}{U} \big| \Delta \big|^2 - a \big| \Delta + t_\text{hyb} \big|^2 + b \big| \Delta + t_\text{hyb} \big|^4
\end{split}\end{align}
in the amplitude-phase representation.
If $\frac{1}{U}$ and $a$ become comparable in size,
the potential gives rise to squeezed equipotential lines similar to the phase space trajectories observed in the main text.
We show one example in Fig.~\ref{suppfig:traj_gl}.

\subsection{Derivation}
The derivation of such a Ginzburg-Landau Lagrangian can be based formally on a derivative expansion within the effective action formalism.
We provide some details of this derivation in the following.

In order to perform the derivative expansion,
we start from the real-time/Minkowski action formalism and perform a Wick rotation to imaginary times later ($t = -i\tau$, $-i \partial_t \leftrightarrow \partial_\tau$).
We use the definitions
\begin{align}\begin{split}
Z = \int D[\Psi,\overline{\Psi}] \text{e}^{i S_\text{M}[ \Psi, \overline{\Psi}]}, \quad S_\text{M} = \int L_\text{M} \text{d}t, \quad L_\text{M} = - \overline{\Psi} (-i\partial_t) \Psi - H
\end{split}\end{align}

We can start from the mean-field Hamiltonian ($\Psi_i = (v_i, c_i)^T$, $\phi = -U \langle v_i^\dagger c_i \rangle$, $n_v = \langle v_i^\dagger v_i \rangle$, $n_c = \langle c_i^\dagger c_i \rangle$) formulated in momentum space:
\begin{align}\begin{split}
 H = \sum_k \Psi_k^\dagger \begin{pmatrix} -\epsilon_k + U \big( n_c - \frac{1}{2} \big) - \frac{\Delta\epsilon}{2} & \phi + t_\text{hyb} \\ \overline{\phi} + t_\text{hyb} & +\epsilon_k + U \big( n_v - \frac{1}{2} \big) + \frac{\Delta\epsilon}{2} \end{pmatrix} \Psi_k + \frac{1}{U} | \phi |^2 + U \Big( n_v n_c + \frac{1}{4} \Big)
 \label{eq:mf_ham}
\end{split}\end{align}
and allow the field $\phi$ to fluctuate such that we arrive at an effective action $Z = \int D[\phi,\overline{\phi}] \prod_k D\big[ \Psi_k, \overline{\Psi}_k \big] \text{e}^{i S[ \Psi_k, \overline{\Psi}_k, \phi]}$, which reads
\begin{align}\begin{split}
 S &= \int_0^\infty \text{d}t\; \Big[ \sum_k \Big( \overline{\Psi}_k \big( i \partial_t \big) \Psi_k - \overline{\Psi}_k T_k \Psi_k \Big) - \frac{1}{U} | \phi |^2 - U \Big( n_v n_c + \frac{1}{4} \Big) \Big]
\end{split}\end{align}
and $T_k$ is the matrix in \eqref{eq:mf_ham}.
In the next step, we would like to integrate out the fermionic fields and need to be careful what happens with the temporal derivative.
For the action $S_0 = \int \big( i \overline{\Psi}(t) \dot\Psi(t) - \overline{\Psi}(t) T \Psi(t) \big) \text{d}t =: \int L_0(t) \text{d}t$ we obtain the kernel (one-loop effective action)
\begin{align}\begin{split}
 \frac{\delta S_0}{\delta \Psi(s)} &= \frac{\partial L_0}{\partial \Psi}(s) - \partial_t \Big( \frac{\partial L_0}{\partial \dot\Psi} \Big)(s) = - \overline{\Psi}(s) T + i \dot{\overline{\Psi}}(s) \\
 \frac{\delta S_0}{\delta\overline{\Psi}(s') \delta\Psi(s)} &= - T \delta(s-s') + i \dot\delta(s-s') = \big( i \partial_s - T \big) \delta(s-s') ,
\end{split}\end{align}
where in the first line we need to be aware of the minus sign that occurs upon moving a Grassmann derivative through a Grassmann variable.
Effectively, the temporal derivative acts on a delta function, which will give rise to one frequency integral.
The action transforms into
\begin{align}\begin{split}
 S_0 &= \int_0^\infty \text{d}t\; \sum_k \operatorname{tr}\ln\big[ \big( i \partial_t - T_k \big) \big] \delta(t-t')|_{t=t'}
\end{split}\end{align}
Inserting the delta function representation $\delta(t) = \frac{1}{2\pi} \int_{-\infty}^\infty \text{e}^{i \omega t} \text{d}\omega$ and shifting the $t'$-part to the left yields:
\begin{align}\begin{split}
 S_0 &= \int_0^\infty \text{d}t\; \int_{-\infty}^\infty \text{d}\omega \sum_k \text{e}^{- i \omega t} \operatorname{tr}\ln\big[ \big( i \partial_t - T_k \big) \big] \text{e}^{i \omega t}
 \label{eq:S_0_integral}
\end{split}\end{align}
If one assumes $T_k$ to be time-independent,
one can evaluate the derivative (will be replaced by $i\omega$) and obtains BCS mean-field theory.
If, however, this is not the case, the matrix operator cannot be simply evaluated but one can do a derivative expansion.
The simplest approach is to expand in the off-diagonals of the Green's function $G = \big( i \partial_t - T_k \big)^{-1}$
using $G^{-1} = G_0^{-1} + X$, where
\begin{align}\begin{split}
 G_0 = \begin{pmatrix} \Big( i\partial_t - \epsilon_k^{(v)} \Big)^{-1} & 0 \\ 0 & \Big( i\partial_t - \epsilon_k^{(c)} \Big)^{-1} \end{pmatrix}, \quad
 X = \begin{pmatrix} 0 & \phi + t_\text{hyb} \\ \overline{\phi} + t_\text{hyb} & 0 \end{pmatrix}
\end{split}\end{align}
with $\epsilon_k^{(v)} = \epsilon_k - U ( n_c - \frac{1}{2} ) + \frac{\Delta\epsilon}{2}$ and $\epsilon_k^{(c)} = -\epsilon_k - U ( n_v - \frac{1}{2} ) - \frac{\Delta\epsilon}{2}$.
At half filling, $n_v + n_c = 1$ and $n_v = \frac{1}{2} ( 1 + \Delta n )$, $n_c = \frac{1}{2} ( 1 - \Delta n )$ for $\Delta n = n_v - n_c$.
Hence $\epsilon_k^{(v)} = \epsilon_k + \frac{U}{2} \Delta n + \Delta\epsilon =: \tilde{\epsilon}_k = -\epsilon_k^{(c)}$.

We get
\begin{align}\begin{split}
 \operatorname{tr}\ln\big[ G_0^{-1} + X \big] &= \operatorname{tr}\ln\big[ G_0^{-1} ( 1 + G_0 X ) \big] = \operatorname{tr}\ln\big[ G_0^{-1} \big] + \operatorname{tr}\ln\big[ 1 + G_0 X \big] \\
 &= \operatorname{tr}\ln\big[ G_0^{-1} \big] - \sum_{n=1}^\infty \frac{1}{2n} \operatorname{tr}\big[ \big( G_0 X \big)^n \big]
\end{split}\end{align}

The second order term has got the structure $\operatorname{tr}\big[ G_0 X G_0 X \big]$,
writing out the matrix multiplications yields ($\phi_h = \phi + t_\text{hyb}$ for simplicity)
\begin{align}\begin{split}
 \operatorname{tr}\big[ G_0 X G_0 X \big] &= (G_0)_{11} \; \phi_h \; (G_0)_{22} \; \overline{\phi_h} \; + \; (G_0)_{22} \; \overline{\phi_h} \; (G_0)_{11} \; \phi_h \\
 &= \; \big( -\omega - \tilde{\epsilon}_k \big)^{-1} \big( -\omega -(i\tilde{\partial_t}) + \tilde{\epsilon}_k \big)^{-1} \; \phi_h \; \overline{\phi_h} \\
 &\;\, + \big( -\omega + \tilde{\epsilon}_k \big)^{-1} \big( -\omega -(i\tilde{\partial_t}) - \tilde{\epsilon}_k \big)^{-1} \; \overline{\phi_h} \; \phi_h 
\end{split}\end{align}
In order to evaluate \eqref{eq:S_0_integral},
we need to take into account the action of the derivative operators in $G_0$ both on $\phi_h$ and on the complex exponentials.
We follow the strategy outlined by Schakel~\cite{Schakel2000} and commute all derivative operators to the left of the operator string as a first step.
One can read the right hand side of the product rule $f \big( \partial_t g \big) = \partial_t \big( f g \big) - \big( \partial_t f \big) g$ as one derivative that acts on everything to its right plus one derivative~($\tilde{\partial_t}$) that only acts to the next object to its right.
The latter gives rise to the derivative expansion of the order parameter,
while the first can be treated using partial integration such that it acts on the $\text{e}^{-i \omega t}$ on the left.

We will now go to imaginary times, both for the fermion frequency $\omega$ as well as for the derivative operator $\tilde{\partial_t}$ that acts on the bosonic field $\Delta$ and therefore transforms into a bosonic Matsubara frequency.
For $\omega$, we use the replacement $\int \frac{\text{d}\omega}{2\pi} g(\omega) \rightarrow \frac{i}{\beta} \sum_{n \in \mathcal{Z}} g(i \omega_n)$ and the following Matsubara summation (for fermionic $\omega_n$ and bosonic $\omega_l$),
\begin{align}\begin{split}
 \frac{1}{\beta} \sum_{n \in \mathcal{Z}} \frac{1}{i\omega_n + i\omega_l - \epsilon} \frac{1}{i\omega_n - \epsilon'} = -\frac{n_\text{F}(\epsilon) - n_\text{F}(\epsilon')}{i\omega_l -\epsilon + \epsilon'}
\end{split}\end{align}
This expression can now be continued (back) analytically for the bosonic Matsubara frequency $i\omega_l$ to $\omega + i 0^+$ and (ultimately)) Fourier-transformed to the real-time derivative operator $i \partial_t$.
In our case, $\epsilon = \tilde{\epsilon}_k$ and $\epsilon' = -\tilde{\epsilon}_k$.
At half filling, $\mu = 0$ and therefore $n_\text{F}(-x) = 1 - n_\text{F}(x)$ such that
\begin{align}\begin{split}
 -\frac{n_\text{F}(\tilde{\epsilon}_k) - n_\text{F}(-\tilde{\epsilon}_k)}{\omega -2\tilde{\epsilon}_k + i 0^+} = \Big( i \pi \delta(\omega -2\tilde{\epsilon}_k) + \mathcal{P} \frac{1}{\omega -2\tilde{\epsilon}_k} \Big) \Big( 1 - 2 n_\text{F}(\tilde{\epsilon}_k) \Big)
\end{split}\end{align}
The sum over crystal momenta can be transformed into an integral over energies, if the density of states $\nu(\epsilon)$ is introduced into the equation: $\displaystyle \sum_k = \int \text{d}\epsilon \; \nu(\epsilon)$:
\begin{align}\begin{split}
 \int \text{d}\tilde{\epsilon} \; \nu(\tilde{\epsilon}) \Big[ -\frac{n_\text{F}(\tilde{\epsilon}) - n_\text{F}(-\tilde{\epsilon})}{\omega -2\tilde{\epsilon} + i 0^+} \Big] &= i \pi \int \text{d}\epsilon \; \nu(\epsilon) \; \delta\Big(\frac{\omega}{2} - \epsilon \Big) \Big( \frac{1}{2} - n_\text{F}(\epsilon) \Big) + \mathcal{P} \int \text{d}\epsilon \; \nu(\epsilon) \; \frac{1 - 2 n_\text{F}(\epsilon)}{\omega -2 \epsilon} \\
 &= i\pi \nu( \omega/2) \Big( \frac{1}{2} - n_\text{F}(\frac{\omega}{2}) \Big) + \mathcal{P} \int \text{d}\epsilon \; \nu(\epsilon) \; \frac{1 - 2 n_\text{F}(\epsilon)}{\omega -2 \epsilon}
\end{split}\end{align}
Let us expand the expression to leading order in $\omega$:
\begin{align}\begin{split}
 i\pi \nu( \omega/2) \Big( \frac{\omega}{8T} + \mathcal{O}(\omega^2) \Big) + \mathcal{P} \int \text{d}\epsilon \; \nu(\epsilon) \; \frac{1 - 2 n_\text{F}(\epsilon)}{ -2 \epsilon} \Big( 1 + \frac{\omega}{2\epsilon} - \frac{\omega^2}{4 \epsilon^2} \Big)
\end{split}\end{align}
$1 - 2 n_\text{F}(\epsilon) = n_\text{F}(\epsilon) - n_\text{F}(-\epsilon)$ is an odd function of $\epsilon$, $\nu(\epsilon)$ is even.
Looking for the principal value around zero, this implies that the principal value of the linear-in-$\omega$ summand ($\sim \epsilon^{-2}$) vanishes.
The other two terms remain with a numerical prefactor.
Fourier-transforming back from $\omega$ to $i \partial_t$ yields kinetic Lagrangian terms of the structure $\overline{\phi_h} (i\partial_t) \phi_h$ and $\overline{\phi_h} (i\partial_t)^2 \phi_h$.

Under the assumption of a constant order parameter amplitude,
the second term yields for $\phi = |\phi| \text{e}^{i\theta}$ a contribution $\overline{\phi} (i\partial_t)^2 \phi = - |\phi|^2 \big( i \ddot\theta - \dot\theta^2 \big)$ giving rise to the term $\dot\theta^2$ in the phase mode Hamiltonian.
Similarly, the potential term $| \phi + t_\text{hyb} |^2 = | \phi |^2 + 2 t_\text{hyb} \operatorname{Re}(\phi) + t_\text{hyb}^2$ hosts the contribution $2 t_\text{hyb} \operatorname{Re}(\phi) = 2 t_\text{hyb} |\phi| \cos(\theta) \equiv \omega_\text{ph} \cos(\theta)$.

\bibliography{supplement.bib}